\documentclass[prl,superscriptaddress,preprint,amsmath,amssymb]{revtex4}
\usepackage{graphicx}
\usepackage{indentfirst}
\usepackage{psfrag}
\usepackage{epsfig}
\usepackage{amsmath}
\usepackage{amssymb}
\usepackage{bm}

\begin{document}
\renewcommand{\baselinestretch}{2}


  \title{Angle-Suppressed Scattering and Optical Forces on Submicron Dielectric Particles}

\author{M. Nieto-Vesperinas, $^{1,*}$, R. Gomez-Medina$^{1,*}$, J. J. Saenz, $^{2,* *}$}

\address{$^1$Instituto de Ciencia de Materiales de Madrid, Consejo Superior
de Investigaciones Cientificas, Campus de Cantoblanco, Madrid 28049,
Spain.
\\
$^2$Departamento de F\'{\i}sica de la Materia Condensada,
Universidad Aut\'{o}noma de Madrid, 28049 Madrid, Spain, and \\
Donostia International Physics Center (DIPC), Paseo Manuel
Lardizabal 4, 20018 Donostia-San Sebastian, Spain.\\
$^*$mnieto@icmm.csic.es \, \, $^{* *}$juanjo.saenz@uam.es}


\begin{abstract}
We  show that submicron Silicon spheres, whose polarizabilities are
completely given by their two first Mie coefficients, are an
excellent laboratory to test effects of both angle-suppressed and
resonant  differential scattering cross sections. Specifically,
outstanding scattering angular distributions, with zero forward or
backward scattered intensity, (i.e., the so-called Kerker's
conditions), previously discussed for hypothetical magnetodielectric
particles, are now observed for those Si objects in the near
infrared. Interesting new consequences for the corresponding optical
forces are derived from the interplay, both in and out resonance,
between the electric and magnetic induced dipoles.
\end{abstract}
\maketitle


\section{Introduction}
The scattering properties of small particles having special
electromagnetic properties has long been a topic of theoretical
interest \cite{123, Bohren}. Even in the simplest case of small or
of dipolar scatteres, remarkable scattering effects of
magnetodielectric particles were theoretically  established by
Kerker \cite{Kerker} concerning suppression or minimization of
either forward or backward scattering. Notwithstanding, no concrete
example of such particles that might present those interesting
scattering properties in the visible or infrared regions has been
proposed yet. Intriguing applications in  scattering cancelation and
cloaking \cite{Leon, Pendry} have renewed the interest in the field
\cite{Alu}.

 The interplay between electric and magnetic properties is a key ingredient determining the scattering
  characteristics of small objects. It also has a key role in the study of magneto-optical systems \cite{Lakhtakia, Farias, Braulio, Silvia}
    or in the quest for magnetic plasmons \cite{Engheta1}. The unavoidable problems of losses and
     saturation effects inherent to metamaterials in the optical and near infrared regimes have stimulated the study
      of high permittivity particles as their constitutive elements
 \cite{Kevin8, Wheeler_2009, Kevin9, Kevin10, Kevin11, Kevin12, Kevin} with unique electromagnetic properties, and antennas based on dielectric
resonators \cite{Mongia, Hsu, Brongersma}. As regards radiation
pressure, Ashkin \cite{Ash_reson} was the first to observe the
effect of both their electric and magnetic resonances, which was
theoretically analyzed by Chylek \cite{Chylek} also in connection
with higher order Mie coefficients. The first order resonances were
subsequently theoretically studied \cite{Videen}

In this work we first address small dielectric particles, described
by the first order Mie coefficients, as regards scattering
properties similar to those reported for magnetodielectric spheres
\cite{Kerker,GCamara}.  Secondly, we analyze how those scattering
effects  affect  the radiation pressure exerted by the
electromagnetic field on such particles \cite{Ashkin2, Ashkin3,
MNVOL,  Grier, Neuman, MNV&Ricardo, MNV1, Ratner,Albala}. This is
relevant in the study of light induced interactions \cite{Golov,
Chaumet, Chaumet_binding, Dholakia} and dynamics \cite{Albaladejo,
Roichman, Albala2, Wu, Zapata} of particles trapped or moved by
light, topics with increasing number of applications.

Only recently, a theory of optical forces on small magnetodielectric
particles has been developed \cite{Chaumet_magnet,MNV&JJ}. This
includes pure dielectric particles which can be  well described by
their  two first electric and magnetic Mie coefficients
\cite{MNV&JJ}; but again no concrete particles were addressed.
However, in a later work, we have shown that Silicon spheres present
dipolar magnetic and electric responses, characterized by their
respective first order Mie coefficient, in the near infrared
\cite{Juanjo}, in such a way that either of them can be selected by
choosing the illumination wavelength. In the present work we show
that they constitute such a previously quested real example of
dipolar particle with either electric and/or magnetic response, of
consequences both for their emitted intensity and behavior under
electromagnetic forces.

This paper is organized as follows: in Section 2 we discuss the
scattering cross section properties of a magnetodielectric particle,
and we propose a generalization of the so-called second Kerker
condition \cite{Kerker}. Then we introduce the real instance of a
small Si sphere that illustrates these characteristics. It should be
stressed that as far as we know, this is the first concrete example
of such a kind of dipolar magnetodielectric particle, from whose
resonances one can observe consequences on both the scattering cross
section and the optical forces at different wavelengths in the near
infrared. In Section 3 we address the optical force on such
particles from an incident plane wave. We obtain an expression for
this force in terms of the differential scattering cross sections
and discuss the consequences, depending on the polarizabilities. In
particular, we study the conditions for a minimum force, as well as
the resulting force when the first and generalized second Kerker
conditions hold. Then we illustrate these forces with the small Si
sphere. The results indicate that this particle  may suffer enhanced
radiation pressure which is mainly due to the resonant induction of
its magnetic dipole.



\section{Scattering cross sections. Kerker conditions}

Let us consider a small sphere of radius $a$ immersed in an
arbitrary lossless medium with  relative dielectric permittivity
$\epsilon$ and magnetic permeability $\mu$.  Under illumination by
an external field of frequency $\omega$, ${\bf E}={\bf E}^{(i)}({\bf
r}) e^{-i\omega t}$, ${\bf B}= {\bf B}^{(i)}({\bf r}) e^{-i\omega t}
$, the induced electric and magnetic dipoles ${\bf p}$ and ${\bf m}$
are written in terms of the  electric and magnetic complex
polarizabilities $\alpha_e$ and $\alpha_m$ as: ${\bf p} = \alpha_e
{\bf E}^{(i)}$ and ${\bf m} = \alpha_m {\bf B}^{(i)}$. For a small
sphere, with constitutive parameters $\epsilon_p$ and $\mu_p$, the
dynamic polarizabilities are expressed in terms of  the Mie
coefficients $a_1$ and $b_1$ as \cite{123} : $
 \alpha_e =  3i \epsilon a_1 /(2 k ^3) $ and
$\alpha_m = 3 i b_1 /(2\mu k ^3), $ ($k$ is the wavenumber: $k =
\sqrt{\epsilon \mu} \ \omega/c$), which may be written in the form
\cite{MNV&JJ}:
\begin{eqnarray}
\alpha_e = \frac{\alpha_e^{(0)}}{ 1- i \frac{2}{3 \epsilon } k ^3
\alpha_e^{(0)}}, \,\,\, \alpha_m = \frac{\alpha_m^{(0)}}{1- i
\frac{2}{3} \mu  k ^3 \alpha_m^{(0)}}. \label{Draine}
\end{eqnarray}
In Eq. (\ref{Draine})
$\alpha_e^{(0)}$ and $\alpha_m^{(0)}$ are static polarizabilities.
The particle extinction, $\sigma^{(\text{ext})}$, absorption,
$\sigma^{(a)}$ and scattering, $\sigma^{(s)}$, cross sections are
written in terms of the polarizabilities as
 \begin{eqnarray}
 \sigma^{(\text{ext})} &=&  4 \pi k \Im\left\{ \epsilon^{-1} \alpha_e + \mu \alpha_m \right\} \label{ext} \\
 \sigma^{(s)} &=& \frac{8\pi}{3} k^4 \left\{ |\epsilon^{-1} \alpha_e|^2 + |\mu \alpha_m|^2 \right\}.
 \end{eqnarray}
The symbol $\Im$ means imaginary part. Energy conservation, i.e. the
so-called "Optical Theorem", Eq. (\ref{ext}), imposes $
\sigma^{(\text{ext})} = \sigma^{(s)} + \sigma^{(a)} $.

In terms of the static polarizabilities,  the absorption cross
section is written as
\begin{eqnarray}
\sigma^{(a)}= 4\pi k [(\epsilon A)^{-1} \Im \alpha_e^{(0)}
+ \mu B^{-1} \Im \alpha_{m}^{(0)} ],  \label{absorpt} \\
A=|1- i \frac{2}{3\epsilon } k ^3 \alpha_e^{(0)}|^2, \,\,\, B=|1- i
\frac{2}{3} \mu k ^3 \alpha_m^{(0)}|^2 , \nonumber
\end{eqnarray}

In absence of magnetic response, i.e. for an induced pure electric
dipole (PED), the far field radiation pattern is given by the
differential scattering cross section which, averaged over incident
polarizations, is
 \cite{Jackson}:
\begin{eqnarray}
 \frac{d\sigma_{\text{PED}}^{(s)}}{d\Omega} (\theta) = \frac{k ^4}{2}
\left| \epsilon^{-1} \alpha_e\right|^2 \left(1+ \cos^2
\theta\right), \label{difcrossped}
\end{eqnarray}
being symmetrically distributed between forward and backward
scattering. However, when we consider the contribution of both the
electric and magnetic induced dipoles, we obtain
 \cite{MNV&JJ, Jackson}:
\begin{eqnarray}
 \frac{d\sigma^{(s)}}{d\Omega} (\theta)= && \frac{k ^4}{2}   \left(
\left| \epsilon^{-1} \alpha_e\right|^2+\left|\mu \alpha_m\right|^2
\right) (1+ \cos^2 \theta)  \nonumber \\ &+&
 2 k^4  \frac{\mu}{\epsilon} \Re(\alpha_e \alpha_m^* ) \cos \theta ,
\label{difcross2}
\end{eqnarray}
which is mainly distributed in the forward or backward region
according to whether $\Re(\alpha_{e} \alpha_{m}^{*})$ is positive or
negative, respectively. The symbol $\Re$ means real part. In
particular, the forward ($\theta = 0^\circ$; "+") and backward
($\theta = 180^\circ $; "-") directions, the intensities are simply
given by
 \begin{eqnarray}
  \frac{d\sigma^{(s)}}{d\Omega}(\pm)  &=& k ^4
\left| \epsilon^{-1} \alpha_e \pm \mu \alpha_m \right|^2.
 \end{eqnarray}
This asymmetry, arising from the interference between the electric
and magnetic dipolar fields, lead to a number of interesting
effects:

{\em i) The intensity in the backscattering direction can be exactly
zero} whenever \begin{eqnarray}
 \epsilon^{-1} \alpha_e = \mu
\alpha_m \quad \Rightarrow \quad
\frac{d\sigma^{(s)}}{d\Omega}(180^\circ) = 0 \label{firstKK}.
\end{eqnarray}
This anomaly corresponds to the so-called {\em first Kerker
condition} \cite{Kerker}, theoretically predicted for
magnetodielectric particles having $\epsilon_p = \mu_p$.

{\em ii)} Although  the intensity cannot be zero in the forward
direction, (causality imposes $\Im\{\alpha_e\}, \Im\{\alpha_m\} >
0$),  {\em  in absence of particle absorption, the forward scattered
intensity is near a minimum} at
\begin{eqnarray}
\Re\{\epsilon^{-1} \alpha_e \} &=& -\Re\{\mu \alpha_m\},  \ \
 \ \Im\{\epsilon^{-1} \alpha_e \} = \Im\{\mu \alpha_m\} \nonumber
\\  \ \ \Rightarrow  \frac{d\sigma^{(s)}}{d\Omega}(0^\circ)   &=&  k^4 \left| 2 \Im\{\epsilon^{-1} \alpha_e\}\right|^2 =
 \frac{16}{9} k^{10} \left|\epsilon^{-1} \alpha_e\right|^4 \nonumber \\ &=&
\left| \frac{2}{3} k^3 \epsilon^{-1}\alpha_e \right|^2
\frac{d\sigma^{(s)}}{d\Omega}(180^\circ). \label{secondKK}
\end{eqnarray}
(Notice that the first line of Eq. (\ref{secondKK}) leads to a
minimum of the intensity if in addition:  $Im\{\epsilon^{-1}
\alpha_e \} = \Im\{\mu \alpha_m\}= minimum$). For lossless
magnetodielectric particles, Eq. (\ref{secondKK}) is known as the
{\em second Kerker condition}, and leads exactly to a zero minimum
of $d\sigma^{(s)}(0^\circ)/d\Omega$ \cite{Kerker,GCamara} when
$\epsilon_p = -(\mu_p-4)/(2\mu_p+1)$ and the particle scattering is
well characterized by the quasistatic approximation: $\Re\alpha
\approx \Re\alpha^{(0)}$, $\Im\alpha \approx \Im\alpha^{(0)}\approx
0$, of the Rayleigh limit: $ka\ll 1$, $k|n_p| a\ll 1$, in which
\cite{MNV&JJ, Bohren}: $\alpha_e^{(0)}=\epsilon a^3
\frac{\epsilon_p-\epsilon }{\epsilon_p+ 2 \epsilon }$,
$\alpha_m^{(0)}=\mu ^{-1} a^3 \frac{\mu_p-\mu }{\mu_p+ 2 \mu }$. As
a matter of fact, this condition was derived in \cite{Kerker} under
these approximations.
 It should be remarked that the actual
intensity for a very small particle goes as $\sim (ka)^{10}$, which
only when $ka$ is well below 1, would be negligible \cite{Alu}.
Otherwise, as is the case of the small particles here addressed,
this intensity is near a non-zero minimum value of
$d\sigma^{(s)}(0^\circ)/d\Omega$, as seen in Section 3. Although
being of fundamental interest, no concrete example of dipolar
magnetodielectric particles that might present such anomalous
scattering in the visible or infrared regions has been proposed.

Our derivation of the special scattering conditions (\ref{firstKK})
and (\ref{secondKK}) was obtained with the unique assumption that
the radiation fields are well described by dipolar electric and
magnetic fields, including their generalization in terms of the
coefficients $a_1$ and $b_1$. This goes well beyond the Rayleigh
limit and should apply to any small particle described by Eqs.
(\ref{Draine}) in terms of these two Mie coefficients. The first
line of Eq. (\ref{secondKK})  can then be considered as a {\em
generalized second Kerker condition}, and is the first result of
this work. Specifically, {\it the second Kerker condition Eq.
 (\ref{secondKK}) also applies to purely dielectric
spheres ($\mu_p =1$) providing that their scattering properties may
be fully described by the two first terms in the Mie expansion}.

\section{An instance of magnetodielectric particle: A Silicon sphere}

A recent work \cite{Juanjo} reports that dielectric spheres whose
refractive index is around 3.5 and have size parameter $ka$ between
0.75 and 1.5
produce a plane wave scattering which is with great accuracy given
by only the two first Mie coefficients $a_1$ and $b_1$, [see Eq.
(\ref{Draine})]. Here we next show that they are very convenient,
real and unexpected objects, for testing Kerker conditions, as well
as new scattering effects and their consequences on optical forces.

An example is a Silicon sphere of radius $a=230 nm$, whose
refractive index may well be approximated by $\epsilon_p=3.5$ in the
range of near infrared wavelengths ($\lambda \approx 1.2-2 \mu$m )
of this study \cite{Juanjo}.


\begin{figure}
\includegraphics[width=8cm]{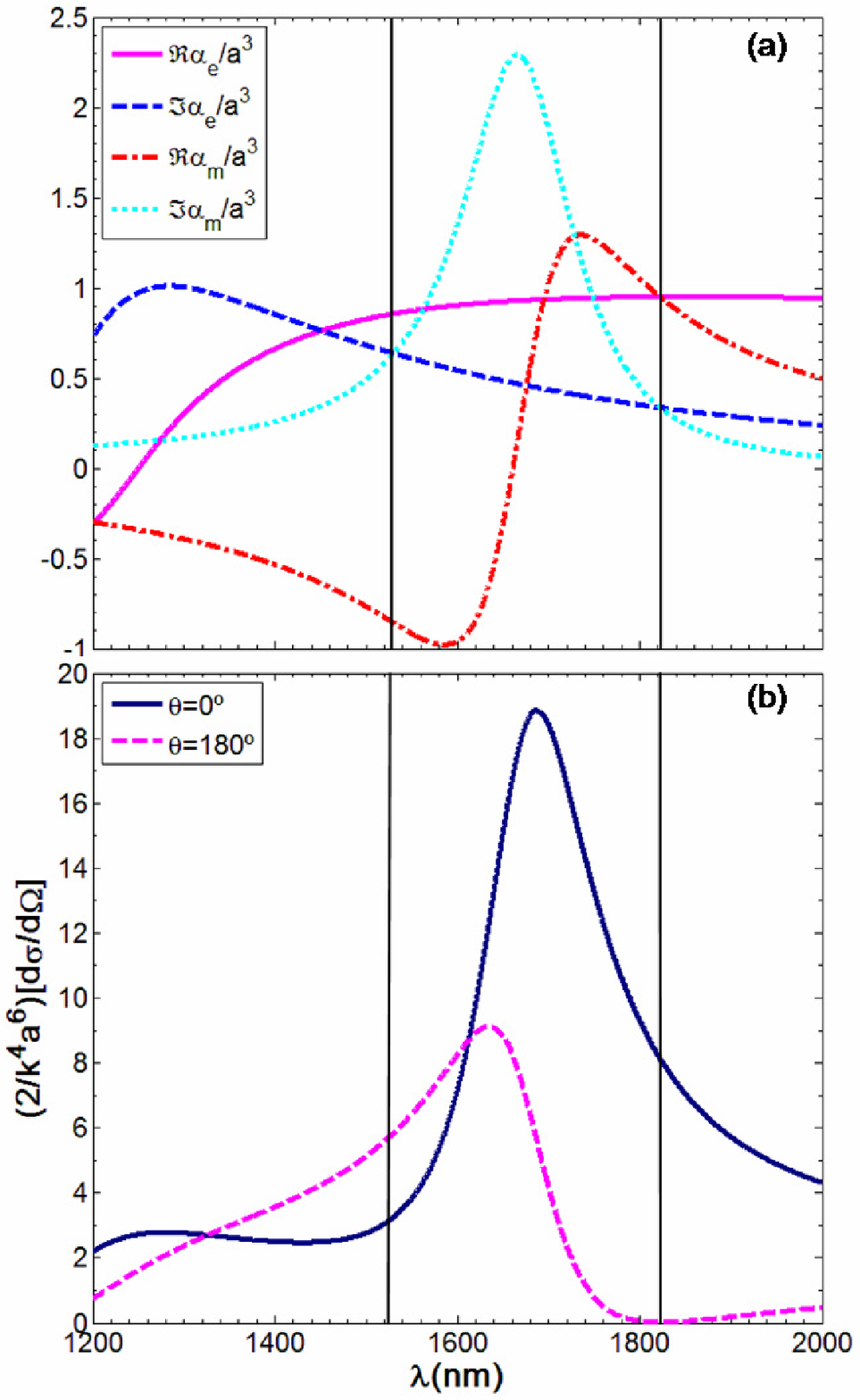}
\caption{Results for a Si sphere of radius $a=230nm$;
$\epsilon_p=12$ and $\mu_p=1$. The host medium has $\epsilon=\mu=1$.
(a) Normalized real and imaginary parts of both the electric and
magnetic polarizabilities. (b) Normalized differential scattering
cross section in the forward and backscattering direction. The first
and second Kerker conditions are marked by the right and left
vertical lines, respectively. }
\end{figure}

Figure 1 (a) shows the real and imaginary parts of the
polarizabilities, [Eq. (\ref{Draine})], whereas Fig. 1 (b) contains
the differential scattering cross sections in the forward and
backscattering directions. The maxima in $\alpha_e$ and $\alpha_m$,
[see Fig. 1(a)], occur around $1300$ nm and $1700$ nm, respectively,
and are well separated from each other. The sharp peaks of the
differential scattering cross sections, [Fig. 1(b)], are mainly due
to the corresponding dominant magnetic dipole contribution
$\alpha_m$ near the first Mie resonance. One sees the values of
$\lambda\simeq 1825$ nm and $1530$ nm at which
$\Im\{\alpha_e\}=\Im\{\alpha_m\}$, which are where the first and
second Kerker conditions hold for these polarizabilities,
respectively.

While the backward intensity drops to zero at the first Kerker
condition wavelength, at the frequency of the second condition the
radiated intensity is near a non-zero minimum in the forward
direction. {\it Dielectric spheres and, in particular, lossless Si
particles in the near infrared,  then constitute a realizable
laboratory to observe such a special scattering}. This is another
main result of the present work.

It should be observed that {\it for a lossless particle as the one
under study, the two Kerker conditions are a consequence of the
optical theorem, Eq.  (\ref{ext}), written for the electric and for
the magnetic dipole, separately}. This, in turn, obeys to the zero
contribution of the self-interaction term between both dipoles,
[i.e., the last term of Eq.(\ref{difcross2})], to the total
scattering cross section. Then, if one imposes the equality of
imaginary parts: $\Im \alpha_e=\Im \alpha_m$, and substracts from
each other the optical theorem equations of each dipole,  one
immediately derives that $\Re \alpha_e=\pm \Re \alpha_m$.

\section{Effects on optical forces}

It is of interest to analyze the consequences of  anomalous
scattering properties in radiation pressure forces. For an incident
plane wave, ${\bf E}^{(i)} = {\bf e}^{(i)} e^{i k {\bf s}_0 \cdot
{\bf r}}$ and ${\bf B}^{(i)} = {\bf b}^{(i)} e^{i k {\bf s}_0  \cdot
{\bf r}}$, the time averaged force on a dipolar particle is written
as the sum of three terms \cite{MNV&JJ}:
\begin{eqnarray}
<{\bf F}> &=& <{\bf F}_{e}> + <{\bf F}_{m}>+ <{\bf F}_{e-m}>
\nonumber
\\ &=& {\bf s}_0 F_0  \left[ \frac{1}{a^3}\Im\left\{ \epsilon^{-1} \alpha_e  + \mu \alpha_m \right\}
- \frac{2k^3}{3a^3}\frac{\mu}{\epsilon}\Re(\alpha_e
\alpha_m^{\ast})\right].  \nonumber \\ \label{fpwsp1}
\end{eqnarray}
where $ F_0 \equiv \epsilon k a^3 |{\bf e}^{(i)}|^2/2  $. The first
two terms,
 $<{\bf F}_{e}>$ and $<{\bf F}_{m}>$, correspond to the forces on the induced pure electric and magnetic dipoles,
  respectively. $<{\bf F}_{e-m}>$, due to the interaction between both dipoles \cite{Chaumet_magnet, MNV&JJ}, is
  related to the asymmetry in the scattered intensity distribution, [cf. the last term in Eq.
  (\ref{difcross2})]
  \cite{MNV&JJ}.
 From Eqs. (\ref{difcross2}) and (\ref{fpwsp1}), one derives for the radiation pressure force
 \begin{eqnarray}
<{\bf F}> =  {\bf s}_0  F_0 \frac{1}{6 k a^3}
 \left[\frac{d\sigma^{(s)}}{d\Omega}(0^{\circ})+3\frac{d\sigma^{(s)}}{d\Omega}(180^{\circ}) +
 \frac{3}{2\pi}\sigma^{(a)}
 \right].
 \label{Fdsigma}
\end{eqnarray}
 Eq.(\ref{Fdsigma}), which is a main result of this work,
emphasizes the dominant role of the backward scattering on radiation
pressure forces. In turn, this is connected to the asymmetry
parameter $<cos(\theta)>$ of the radiation pressure \cite{123,
Bohren}.  Notice that Eq. (\ref{Fdsigma}), which is also valid for a
pure dipole, either electric or magnetic, shows that the force due
to a plane wave, which is all radiation pressure \cite{MNV&JJ},
cannot be negative for ordinary host media with $\epsilon$ and $\mu$
real and positive. This expression also manifests that the weight of
the intensity in the backscattering direction is three times that of
the forward scattered power.

Equations (\ref{fpwsp1}) and (\ref{Fdsigma}) provide an appropriate
framework to discuss the interplay between special scattering
properties and radiation pressure forces. Let us consider as a
reference the standard {\it pure electric dipole} (PED) case in
absence of absorption, on which the force from the plane wave is
\begin{eqnarray}
<{\bf F}>_{\text{PED}}=<{\bf F}_e> = F_0
\frac{2k^3}{3a^3} {\bf s}_{0}|\epsilon^{-1} \alpha_e|^2.
\label{Fdippur}
\end{eqnarray}
(The following arguments would equally apply
with a pure magnetic dipole). At a fixed electric polarizability,
the
 addition of an extra magnetic dipole always leads to an increase of the total cross section. However, it
does not necessarily imply an increase of the total force. 

\subsection{A minimum force}
As shown by Eq. (\ref{fpwsp1}), $<{\bf F}> $ cannot be zero, even if
$\sigma^{(a)}=0$; however, if the particle is lossless, by
expressing the bracket of Eq. (\ref{fpwsp1}) as a hypersurface of
the four variables $\Re \alpha_e$,  $\Im \alpha_e$, $\Re \alpha_m$
and  $\Im \alpha_m$, ($\Im \alpha_e >0$,  $\Im \alpha_m> 0$),  it
has the absolute  minimum when $\epsilon^{-1}\Re \alpha_e = \mu \Re
\alpha_m=0$ which is trivial, of course.

Nevertheless, the section of the surface Eq. (\ref{fpwsp1}) at the
planes $\Re \alpha_e =constant$ and    $\Im \alpha_e =constant$, has
minima at $\mu \Re \alpha_m = (1/2) \epsilon^{-1}\Re \alpha_e$ and
$\mu \Im \alpha_m =(1/2)\epsilon^{-1} \Im \alpha_e$. Then,  Eq.
(\ref{fpwsp1}) shows that this {\it minimum} force is
\begin{eqnarray}
<{\bf F}>=F_0 \frac{2k^3}{3a^3} {\bf s}_{0}
\frac{3}{4}[\frac{3\sigma^{(a)}}{2\pi k^4}+|\epsilon^{-1}
\alpha_e|^2], \label{Fmin}
\end{eqnarray}
which for a lossless particle is $3/4$ that on a pure electric
dipole, Eq. (\ref{Fdippur}). Namely, $<{\bf F}> = \frac{3}{4} <{\bf
F}_e>$ .

(Reciprocally occurs by choosing similar plane cuts for the magnetic
polarizability, then an analogous result is obtained with respect to
a pure magnetic dipole with the minimum force: $F_0 (2k^3)/(3a^3)
{\bf s}_{0} (3/4)[3\sigma^{(a)}/(2\pi k^4)+
\mu^{2}|\alpha_{m}|^{2}]$ when $\epsilon^{-1} \alpha_{e}=(1/2)\mu
\alpha_{m}$).

Also, Eq. (\ref{difcross2})  shows that now the differential
scattering cross section of this magnetodielectric particle is
\begin{equation}
\frac{d\sigma}{d
\Omega}=\frac{k^4}{\epsilon^2}|\alpha_{e}|^{2}[\frac{5}{8}(1+\cos^2
\theta)+ \cos\theta]. \label{dsigmamin}
\end{equation}

\subsection{A generalization of the case of a perfectly conducting
sphere}
 On the other hand, let us consider the case in which $\mu
\alpha_{m} =(-1/2)\epsilon^{-1} \alpha_{e}^{\ast}$.
Then, from Eqs. (\ref{fpwsp1}) and (\ref{difcross2}) one has for the
force on the particle:
\begin{equation}
<{\bf F}>=F_0 \frac{2k^3}{3a^3}{\bf s}_0 \epsilon^{-2}[\frac{3}{4}
|\alpha_{e}|^{2}+(\Re\alpha_e)^2], \label{Fmedio}
\end{equation}
and for the corresponding scattering cross section:
\begin{equation}
\frac{d\sigma}{d
\Omega}=\frac{k^4}{\epsilon^2}[\frac{5}{8}|\alpha_{e}|^{2}(1+\cos^2\theta)-
[(\Re \alpha_e)^2-(\Im \alpha_e)^2] \cos\theta ].
\label{dsigmamedio}
\end{equation}

Equations (\ref{Fmedio}) and (\ref{dsigmamedio}) become for a
non-absorbing Rayleigh particle, for which $\Re \alpha_e \simeq \Re
\alpha_e^{(0)}$  and $\Im \alpha_e \simeq 2/(3\epsilon)k^3
|\alpha_e^{(0)}|^2$:
\begin{equation}
<{\bf F}>=F_0 \frac{2k^3}{3a^3}{\bf s}_0
|\epsilon^{-1}\alpha_{e}^{(0)}|^{2} ; \label{Fperf}
\end{equation}
and:
\begin{equation}
\frac{d\sigma}{d
\Omega}=\frac{k^4}{\epsilon^2}|\alpha_{e}^{(0)}|^{2}[\frac{5}{8}(1+\cos^2
\theta)- \cos\theta]. \label{dsigmaperf}
\end{equation}
Equations (\ref{Fperf}) and (\ref{dsigmaperf}) represent a
generalization of the force and differential scattering cross
section, respectively, that apply to a perfectly conducting sphere
at large wavelengths \cite{MNV&JJ}, for which $\mu
\alpha_{m}^{(0)}=(-1/2)\epsilon^{-1} \alpha_{e}^{(0)} \simeq
(-1/2)a^3$ which is a particular case of the aforementioned
condition: $\mu \alpha_{m} =(-1/2)\epsilon^{-1} \alpha_{e}^{\ast}$.

In addition,  the Rayleigh limit, Eq. (\ref{Fperf}), of  Eq.
(\ref{Fmedio}) turns out  to be $7/4$ the force on a lossless PED,
Eq. (\ref{Fdippur}), when in Eq. (\ref{Fdippur}) one also takes this
Rayleigh limit. (In Eq. (\ref{Fdippur}) the term $ F_0 {\bf s}_0
\sigma^{(a)}/(8\pi k a^3)$ should be added if the particle is
absorbing). Analogously happens for a magnetic dipole, if the
electric polarizability is eliminated instead.

 Notice, however, that since the contribution of the term $\Re(\alpha_ e \alpha_m^{\ast})$
 integrated over $\Omega$ is zero, both differential cross sections, Eqs. (\ref{dsigmamin})
 and (\ref{dsigmamedio}), yield the same total scattering cross section and, hence, the same
radiation pressure excluding the   component of the self-interaction
force $<{\bf F}_{e-m}>$. (Similar arguments hold for a magnetic
dipole by choosing the force hypersurface cut:  $\Re \alpha_m
=constant$ and $\Im \alpha_m =constant$). Thus we have the
interesting result on {\it two particles with the same total
scattering cross section, but quite different differential
scattering cross sections}, in particular in the forward and
backscattering directions, {\it and suffering completely different
forces: the former a force Eq. (\ref{Fmedio}) which in the Rayleigh
limit becomes $7/4$ that of a pure non-absorbing dipole, while the
latter experiencing a minimum force Eq. (\ref{Fmin}) which becomes
$3/4$ that of a pure lossless dipole}.

\subsection{Other relative minimum forces. Kerker conditions}

Another minimum force is obtained from Eq. (\ref{fpwsp1}) under the
condition that $| \epsilon^{-1} \alpha_e |^2$ and  $ | \mu \alpha_m
|^2$ be kept constant. This obviously happens when $\Re(
\epsilon^{-1} \alpha_e \mu \alpha_m^{\ast})=| \epsilon^{-1} \alpha_e
| | \mu \alpha_m |$; then if for instance one keeps the condition: $
| \mu \alpha_m | =(1/2) | \epsilon^{-1} \alpha_e |$, this force
becomes again $3/4$ that of a pure dipole; whereas the differential
scattering cross section of such particle is $9/2$ and $1/2$  that
of a pure dipole in the forward ad backscattering directions,
respectively. This is perfectly explained by Eq.(\ref{Fdsigma})


 On the other hand, when $| \epsilon^{-1} \alpha_e |^2= | \mu \alpha_m |^2$, then Eqs.
(\ref{fpwsp1}) and (\ref{Fdsigma}) show that this minimum force is
equal to that of a pure electric dipole Eq. (\ref{Fdippur}). The
differential scattering cross section Eq. (\ref{difcross2}) then is
zero in the backscattering direction, but is maximum and equal to
four times that of a pure dipole, in the forward direction.
Analogously can be reasoned, as before, with respect to a pure
magnetic dipole if the magnetic parameters are chosen instead.

An important case when $| \epsilon^{-1} \alpha_e |^2= | \mu \alpha_m
|^2$,
 is that in which $\epsilon_p/ \epsilon =
\mu \mu_p$, which implies that $\epsilon^{-1} \alpha_{e}=\mu
\alpha_{m}$, namely, at the {\it first Kerker condition},
Eq.(\ref{firstKK}), then {\it the corresponding force $<{\bf
F}>_{FK}$ that one obtains from Eqs. (\ref{fpwsp1}) and
(\ref{Fdsigma}) is}, eliminating the magnetic constants, {\it
exactly equal to the force on a pure electric dipole} Eq.
(\ref{Fdippur}). Then $<{\bf F}>_{FK} =  <{\bf F}_e>$. (It should be
remarked, however, that in this expression for  $<{\bf F}>_{FK}$,
the term $ F_0 {\bf s}_0 \sigma^{(a)}/(4\pi k a^3)$ should now be
added to Eq. (\ref{Fdippur}) if the particle is absorbing). Or a
reciprocal expression for $<{\bf F}>_{FK}$ in terms of the magnetic
polarizability if one substitutes $\epsilon^{-1}\alpha_{e}$ by
$\mu\alpha_{m}$.

Thus, the only difference between both forces: $<{\bf F}>_{FK}$ on a
particle holding the first Kerker condition and that on a pure
electric dipole $<{\bf F}>_{PED}$, occurs when the particle is
absorbing, then being: $ F_0 {\bf s}_0 \sigma^{(a)}/(8\pi k a^3)$.
An equivalent result appears for a magnetic dipole. Also, Eq.
(\ref{difcross2}) shows that this pure dipole cross section is
non-zero in the backscattering direction, but in the forward
direction it is  $1/4$ of the cross section from a magnetodielectric
particle satisfying the first Kerker condition.

At the {\it second Kerker condition}, Eq. (\ref{secondKK}):
$\epsilon^{-1} \alpha_{e}^{(0) }=- \mu \alpha_{m}^{(0)}$ for
Rayleigh lossless particles in the quasistatic approximation:
$|\alpha_{e}|^2 \simeq |\alpha_{e}^{(0)}|^2$, $\Im \alpha_{e} \simeq
\Im \alpha_{e}^{(0) }=0$, so that $d\sigma (0^{\circ})/d \Omega=0$
and Eq. (\ref{fpwsp1}) should lead to a force:
\begin{equation}
<{\bf F}>_{SK} =F_0 \frac{2k^3}{a^3}{\bf s}_0
|\epsilon^{-1}\alpha_{e}^{(0)}|^{2},  \label{FKerker2}
\end{equation}
which would be {\it three times} that on a lossless pure (electric)
dipole, Eq. (\ref{Fdippur}), in that quasistatic approximation; (as
before, we reciprocally argument in terms of a magnetic dipole if
$\alpha_{m}^{(0)}$ is chosen instead). Hence, the larger weight of
the backscattering cross section in the force, [which in this case
is: $4k^4 (\alpha_{e}^{(0)}/\epsilon)^2$] , manifested by Eq.
(\ref{Fdsigma}), would contribute in this situation to such an
increase of the averaged force on this particle on comparison with
that on a pure dipole.


At the {\it generalized second Kerker condition}, Eq.
(\ref{secondKK}), beyond the Rayleigh limit, the  corresponding
force on the lossless particle then is:
\begin{equation}
<{\bf F}>_{SK} = F_0 \frac{2k^3}{3a^3}{\bf s}_0 \epsilon^{-2}[
|\alpha_{e}|^{2}+2(\Re \alpha_e)^2], \label{FKerker2exact}
\end{equation}
which in the Rayleigh limit: $\Re \alpha_e \simeq \Re
\alpha_e^{(0)}$, $\Im \alpha_e \simeq 2/(3\epsilon)k^3
|\alpha_e^{(0)}|^2$, would become smaller and approximately equal to
the quasistatic value Eq. (\ref{FKerker2}). More generally, when
$|\alpha_{e}|^{2}\simeq (\Re \alpha_e)^2$, Eq. (\ref{FKerker2exact})
would again be three times the force on a pure electric dipole Eq.
(\ref{Fdippur}).

\subsection{Summary of the relationships between forces on different
small spheres and that on a pure dipole}

To summarize this, we conclude that at the first generalized Kerker
condition, Eq. (\ref{firstKK}), the interference term of Eq.
(\ref{fpwsp1}) cancels out the magnetic contribution and we obtain
$<{\bf F}> = <{\bf F}_e>$. At the second Kerker condition, Eq.
(\ref{secondKK}), where the backscattering is enhanced, $<{\bf F}> =
3 <{\bf F}_e>$. {\em Notice  that at both Kerker conditions the
total scattering cross section is exactly the same; although the
radiation pressures differ by a factor of 3}. These properties are
illustrated in Fig. 2, where we show the different contributions to
the total time averaged force on a submicron Si particle.


One can conclude from the above discussion derived from Eqs.
(\ref{fpwsp1}) and (\ref{Fdsigma}), that {\it the force on the
magnetodielectric particle is} near (and {\it equal to} for a
Rayleigh particle) $R$ {\it times that on a pure electric dipole,
($R$ being a real number equal or larger than $3/4$), }, whenever:
\begin{eqnarray}
\mu \Re \alpha_{m} =(1/2)(1 \pm \sqrt{4R-3})\epsilon^{-1} \Re
\alpha_{e}, \nonumber \\
\mu \Im \alpha_{m} =(1/2)|1 \pm \sqrt{4R-3}|\epsilon^{-1} \Im
\alpha_{e}. \label{Nroot}
\end{eqnarray}
Analogously occurs with a pure magnetic dipole whenever
$\epsilon^{-1}\Re \alpha_{e}$ and $\epsilon^{-1}\Im \alpha_{e}$ are
reciprocally  replaced by $\mu \Re \alpha_{m}$ and $\mu \Im
\alpha_{m}$, respectively. Equation (\ref{Nroot}) summarizes the
cases discussed before and shows that $R$ cannot be smaller than $R=
3/4$, which would correspond to the minimum force Eq. (\ref{Fmin}).
The case of the PED corresponds to $R=1$ and the square root in Eq.
(\ref{Nroot}) with negative sign, whereas $R=1$ and the plus sign in
that square root leads to the first Kerker condition. On the other
hand, the case of the generalized second Kerker condition
corresponds to $R=3$ and the minus sign in front of the square root
of Eq.(\ref{Nroot}).

\section{Silicon and other high refractive index dielectric spheres: A laboratory to test optical
forces}

\begin{figure}
\includegraphics[width=8cm]{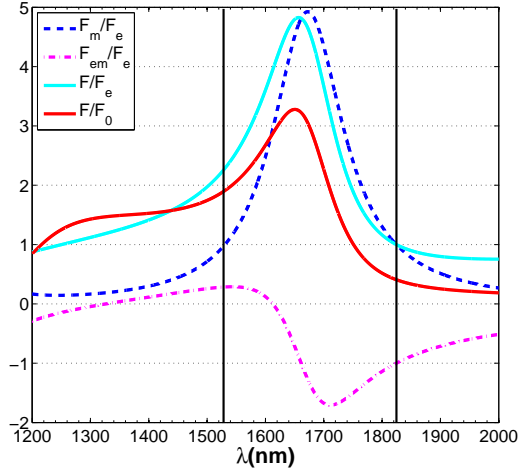}
\caption{Different contributions to the total radiation pressure,
versus the wavelenght, for the Si particle of  Fig. 1. Normalization
is done by either the electric force magnitude $\langle F_e \rangle$
or $F_0=ka^3|{\bf e}^{(i)}|^2/2$ . Again, the vertical lines mark,
from right to left, the first and second Kerker conditions. Notice
that when the first Kerker condition is fullfilled, i.e.
$\Im{\alpha_e}=\Im{\alpha_m}$ and $\Re{\alpha_e}=\Re{\alpha_m}$,
$\langle F \rangle = \langle F_e \rangle = \langle F_m \rangle  = -
\langle F_{e-m} \rangle $. } \label{force}
\end{figure}

Figure 2 shows the different contributions to the total time
averaged force on the Si particle studied in Fig. 1,  presenting
their peaks in the region of wavelengths where the magnetic dipole
dominates. Hence we have the two  additional remarkable results of
this work as follows:

First, {\it there are regions of the spectrum, near the
corresponding electric and magnetic Mie resonances, were
$\Im\alpha_e\gg\Re \alpha_e$ and $\Im\alpha_m\gg\Re \alpha_m$. This
should be observed in future experiments in contrast with previous
observations indicating the opposite result out of resonance}
\cite{Quidant, Righini}. [Notice that Eqs. (\ref{Draine}) show that
at the resonant values of the static polarizabilities
$\alpha_e^{(0)}$ and $\alpha_m^{(0)}$, one has $\Re \alpha_e = \Re
\alpha_m=0$ and $\Im \alpha_e = 3\epsilon/(2k^3)$, $\Im \alpha_m =
3/(\mu 2k^3)$].

Second, {\it the strong peak in the radiation pressure force is
mainly dominated by the first ``magnetic'' Mie resonance, concretely
of $\Im\alpha_m$.  This constitutes an illustration of dipolar
dielectric particle on which the optical force is not solely
described by the electric polarizability}. Also, in such a case the
imaginary part of the polarizability is much larger than its real
part. As a matter of fact, this is the opposite situation to the
usual experiments with optical tweezers out of resonace, in which
gradient forces, (that are proportional to $\Re\{\alpha_e\}$),
dominate over the radiation pressure or scattering force
contribution, (which is proportional to $\Im\{\alpha_e\}$)
\cite{Quidant, Righini}.

Nonetheless, as the size of the particle increases, and {\em for any
dielectric particle},  there is a crossover from electric to
magnetic response as we approach  the first Mie resonance, point at
which there dominance of the magnetic dipole.

Moreover,  just at the resonance, and in absence of absorption,
$\Re\{\alpha_m\} =0$ and $\Im\{\alpha_m\} = 3/(\mu 2k^3)$.
 Then, the radiation pressure contribution of the magnetic term dominates the total force
 $<{\bf F}> \simeq <{\bf F}_m> \approx  (3F_0 {\bf s}_0)/(2k^3 a^3) $. Namely,
 {\em in resonance the radiation pressure force presents a strong peak,  the maximum force being independent of both
 material parameters and particle radius.} On the other hand, the relationship between polarizabilities leading
to Eq. (\ref{Fmedio}), approximately appears in Figs. 1(a) and 1(b)
in the zone about $\lambda\simeq 1450 nm$.

In addition, we observe in Fig. 2 that at the wavelength where the
first Kerker condition holds, as expected from Eq. ({\ref{fpwsp1}}),
the three components of the force are of equal magnitude, but the
electric-magnetic dipole interaction force $<{\bf F}_{e-m}>$
contributes with negative sign and hence the total force equals
either the electric or magnetic contribution, confirming the
previous remarks.  On the other hand, at the wavelength where the
generalized second Kerker condition is fulfilled, the electric and
magnetic force components are equal and the total force, in
agreement wit Eq. (\ref{FKerker2exact}), is almost three times
either of them.

\section{Conclusions}

We have analyzed the scattering properties of magnetodielectric
small particles, proposing a generalization of the second Kerker
condition, and discussed the consequences for the optical forces. We
have shown that real small dielectric particles made of non-magnetic
materials present  scattering properties similar to those previously
reported for somewhat hypothetical magnetodielectric particles
\cite{Kerker}, resulting from an interplay between real and
imaginary parts of both electric and magnetic polarizabilities. Then
we have discussed how these scattering effects do also affect the
radiation pressure on these small particles. Specifically, submicron
Si (as well as Ge and $TiO_2$) particles constitute an excellent
laboratory to observe such remarkable scattering phenomena and force
effects in the near infrared region. This kind of scattering, will
strongly affect the dynamics of particle confinement in optical
traps, which is also governed by both the gradient and curl forces
\cite {MNV&JJ}; and which should be observable as soon as one
introduces a spatial distribution of intensity in the incident
wavefield, and plays with its polarization. We do believe,
therefore, that our results should stimulate further experimental
and theoretical work in this direction, since they suggest
intriguing possibilities in rapid developing fields, ranging from
optical trapping and particle manipulation to cloacking and the
design of optical metamaterials based on lossless dielectric
particles.

\section*{Acknowledgments}
Work supported by the Spanish MEC through the Consolider
\textit{NanoLight} (CSD2007-00046) and FIS2006-11170-C01-C02
 and FIS2009-13430-C01-C02 research grants. RG-M thanks a Juan de
 la Cierva fellowship.

\end{document}